\documentclass[aps,prl,reprint,twocolumn,nofootinbib]{revtex4-2}

\usepackage{epsfig}
\usepackage{amssymb}
\usepackage{amsfonts}
\usepackage{amsbsy}
\usepackage[all,v2,cmtip,2cell]{xy}
\usepackage{amsmath}
\usepackage{dsfont}
\usepackage{bbm}
\usepackage{upgreek}
\usepackage{amssymb,amscd}
\usepackage{graphicx}
\usepackage{mathrsfs}
\usepackage{amsmath,amsthm}
\usepackage{slashed}
\usepackage{dsfont}
\usepackage{tikz}
\usepackage[utf8]{inputenc}

\def\be{ \begin{equation} }
\def\ee{ \end{equation}}

\newcommand{\eq}[1]{\begin{align}\begin{split}#1\end{split}\end{align}}

\def\log{{\rm log}}

\def\CL {{\cal L}}

\def\CS {{\cal S}}

\def\IR{{\mathbb{R}}}

\def\IZ{{\mathbb{Z}}}

\DeclareMathAlphabet{\mathpzc}{OT1}{pzc}{m}{it}

\begin{document}

\title{\boldmath Revisiting 
Scattering Enhancement\\ from the Aharonov-Bohm Effect}

\author{T.~Daniel Brennan, Jaipratap Singh Grewal, and Eric Y. Yang}
\affiliation{Department of Physics, University of California San Diego,\\
 \textit{9500 Gilman Drive, La Jolla CA 92093-0319, USA}}

\begin{abstract}
\noindent 
We revisit the problem of a charged particle scattering off of an Aharonov-Bohm cosmic string. A classic computation gave an infinite total scattering cross section, leading to a Callan-Rubakov-like enhancement which can have important implications on baryon number asymmetry in the early universe. However, unlike the Callan-Rubakov effect, the Aharonov-Bohm interaction is topological and thus it is surprising that it leads to such a dramatic dynamical effect for single particle scattering. We reexamine this old problem through the modern lens of generalized global symmetries by embedding Aharanov-Bohm strings in a discrete gauge theory. We show that the scattering cross section is suppressed by the core size and there is thus no Callan-Rubakov-like enhancement. 
\end{abstract}
\maketitle

\section{Introduction}
\vspace{-0.4cm}

The interaction of dynamical matter with defects is important in the study of QFT given the varied roles defects can play. For example, defect operators can serve as order parameters such as how the expectation value of the `t Hooft operator can detect confinement \cite{tHooft:1977nqb}  or have big phenomenological implications such as through the Callan-Rubakov effect in the interaction of monopole lines with matter 
\cite{Callan:1982au,Callan:1982ah,
Rubakov:1982fp,Polchinski:1984uw,Kitano:2021pwt,Csaki:2021ozp,Hamada:2022eiv,Brennan:2023tae,Brennan:2021ewu,vanBeest:2023dbu,Csaki:2022qtz,Brennan:2024sth}. 


One important class of matter-defect interactions is fermions interacting with cosmic strings. Cosmic strings play a significant part in the study of the early universe: they are  characteristic of axionic dark matter, can lead to important effects on baryogenesis \cite{Brandenberger:1988rp,Brandenberger:1988as,Cline:1998rc,Espinosa:1999jm,Alford:1989ie,Koren:2022axd}, and commonly arise in string theory models \cite{Copeland:2003bj,Polchinski:2005bg}. 

In this paper we will study cosmic strings that interact with free fermions and scalars only via an induced Aharonov-Bohm phase. The case of charged fermions was studied in the classic paper \cite{Alford:1988sj}, where the authors computed a $2d$ differential scattering cross section in the  IR effective theory. They found that the long range Aharonov-Bohm interaction contributes a term to the scattering cross section that is surprisingly unsuppressed by the 
cosmic string core size (similar to the Callan-Rubakov effect). Notably, this leads to an infinite total scattering cross section which can either wash out baryon asymmetry  \cite{Brandenberger:1988rp,Brandenberger:1988as} or be utilized as a mechanism to enhance baryon number violation in the early universe \cite{Alford:1989ie,Koren:2022axd}. 

However, the Callan-Rubakov and Aharonov-Bohm effects are of very different origins: the former involves a dynamical gauge field whereas the latter is mediated by a topological gauge field. In the Callan-Rubakov effect, Dirac fermions couple to a $U(1)$ gauge field and the monopole operator sources a magnetic field that allows chiral symmetries to be violated by the dynamically fluctuating  gauge field via the  ABJ anomaly.  In GUT completions of the Standard Model, this leads to large cross sections for baryon number violating processes. Conversely, in the IR the Aharonov-Bohm string can be described as a surface operator in a $4d$ $\IZ_N$ gauge theory whose interaction with fermions is induced by gauging a $\IZ_N$ subset of a $U(1)$ vector-like symmetry. This operator also activates the ``magnetic'' component of an ABJ anomaly, however, the fact that the $\IZ_N$ gauge field does not have any propagating degrees of freedom suggests that global symmetry violation cannot be activated perturbatively as in the Callan-Rubakov case.

Another puzzling feature of the large scattering cross section is that crossing symmetry relates it to large particle production rates by the cosmic string due to the long range topological interaction. This implies that the  vacuum is unstable around the cosmic string (i.e. a $\IZ_N$-gauge theory surface operator) coupled to charged matter, suggesting that the string is either screened or destabilized. However, this option is not allowed by arguments using generalized symmetry.

Generalized symmetries are the extension of ordinary symmetry groups (which act on local operators/fields) to include the action of all topological operators acting on the set of all operators in a theory (for a review see \cite{Schafer-Nameki:2023jdn,Shao:2023gho,Bhardwaj:2023kri,Brennan:2023mmt}). Generalized symmetries are described by a symmetry category and can protect non-local defect operators such as cosmic strings from decay and play an important role in the dynamics of theories with non-local operators such as in axion-Yang-Mills \cite{Brennan:2020ehu,Brennan:2023kpw,Choi:2022fgx}. For the case of charged matter coupled to Aharonov-Bohm strings, the cosmic string is protected  by a 2-form global symmetry, meaning that the string cannot decay and its charge cannot be screened.

In this paper, we revisit the scattering of charged matter  off of Aharonov-Bohm strings in the IR effective theory to resolve the contradiction between the scattering cross section computation of \cite{Alford:1988sj} and results from generalized symmetries. We show that in the  IR limit (i.e. $\IZ_N$ gauge theory), the cosmic string operator causes the dynamical charged fields to be multi-valued 
which, when taken properly into account, leads to a \emph{vanishing differential cross section} up to corrections suppressed by the core size. Physically, this is because the Aharonov-Bohm effect is topological and does not generate a force that acts on charged matter.

\section{Discrete Gauge Theory and the Aharonov-Bohm Effect}
\vspace{-0.4cm}
We focus on cosmic strings that induce an Aharonov-Bohm phase in units of $2\pi/N$ where $N\in \IZ$. These strings naturally appear in the low energy limit of the charge $N$ Abelian Higgs model which is a $U(1)$ gauge theory with a charge $N$ complex scalar field $\phi$ with potential: 
\eq{
\CL_{\text{UV}}=\frac{1}{2e^2}|dA_1|^2 +|D\phi|^2 -V(\phi)
}
where $D = d + iNA_1$, with $A_1$ a 1-form $U(1)$ gauge field. 

One can then choose $V(\Phi)$ so that the scalar field condenses and the dynamical gauge field is mostly Higgsed, leaving behind a $\IZ_N$ gauge theory described by the action \cite{Banks:2010zn}
\eq{
S_{\text{IR}}=\frac{N}{2\pi}\int dA_1\wedge B_2
}
where $B_2$ is a 2-form $U(1)$ gauge field whose gauge variation is given by $\delta B_2=d\Lambda_1$, with $\Lambda_1$ a $U(1)$-valued 1-form gauge transformation parameter. Note that the equations of motion prevent dynamical propagation of $A_1$ or $B_2$ and the path integral acts like a Fourier transform that localizes to the $\IZ_N$ holonomies of $A_1$ and $B_2$. 

The effective theory admits a pair of operators: Wilson loops and surface/cosmic string operators
\eq{
W_n(\gamma)=e^{i n \oint_\gamma A_1}\quad, \quad \CS_q(\Gamma)=e^{ iq\oint_\Gamma B_2}
}
where $n,q\in 0,...,N-1$. 
Both these operators are topological due to the equations of motion and have mutual winding
\eq{\label{linking}
\left\langle W_n(\gamma)\,\CS_q(\Gamma)\right\rangle=e^{\frac{2\pi i \,nq}{N}{\rm Link}(\gamma,\Gamma)} 
}
where ${\rm Link}(\gamma,\Gamma)$ is the linking number of $\gamma$ with $\Gamma$. 
This linking follows from the fact that inserting the surface operator $\CS_q(\Gamma)$ modifies the equations of motion:
\eq{
dA_1=\frac{2\pi q}{N}\delta(\Gamma)
}
so that $A_1$ has a non-trivial holonomy $\frac{2\pi q}{N}$ around $\Gamma$ which is measured by $W_n(\gamma)$ in the correlation function in \eqref{linking}. In terms of the UV physics, the $B_2$ surface operator carries UV magnetic flux $\Phi=\frac{2\pi q}{eN} $ through its core  (which can heuristically be thought of as being sourced by the rotating phase of $\phi$) -- this magnetic flux  is measured by the holonomy of the $A_1$ Wilson line via Stokes's Theorem. 

In the modern language of generalized symmetries, the Wilson lines are protected  by a $1$-form $\IZ_N^{(1)}$ global symmetry whose action is generated by linking with the $\CS_q$ operators and conversely the $\CS_q$ are protected by a $2$-form $\IZ_N^{(2)}$ global symmetry whose action is generated by linking with the $W_n$. For more details, see \cite{Brennan:2023mmt} and references therein.  

The cosmic strings can be coupled to  a Dirac fermion $\Psi$ of charge $1$ by gauging a $\IZ_N$ subgroup of some $U(1)$ vector-like symmetry group
\eq{\label{LagFerm}
\CL=\frac{N}{2\pi}dA_1\wedge B_2+ \bar\Psi ( i\slashed{D} - m)\Psi~ 
}
which modifies the equations of motion into
\eq{
\frac{N}{2\pi}dA_1=0\quad, \quad \frac{N}{2\pi}dB_2+ \ast j_\Psi=0
}
where $j_\Psi^\mu=i \bar\Psi\gamma^\mu \Psi$ 
so that $\CS_q$ is no longer topological and the $\IZ_N^{(1)}$ global symmetry is explicitly broken. Note that this is consistent with the fact that the Wilson line can be ended on charged local operators.
However, since $dA_1=0$, the Wilson loop is still topological. Although the Wilson loop can be cut, the end points of the Wilson line are not topological, meaning the topological Wilson loop can still enact the $\IZ_N^{(2)}$ global symmetry 
and as such the cosmic string is topologically protected.

Now consider the fermion field in the presence of an $\CS_q(\Sigma)$ operator wrapping the $z$-axis on $\IR^3\times \IR_t$. 
In the partition function, 
$A_1$ is a $U(1)$ gauge field with $\oint \frac{dA_1}{2\pi}\in \IZ$ and $\Psi$ is a single-valued field. After inserting $\CS_q$ into the path integral, the $B_2$ field can be integrated out by shifting $A_1\mapsto A_1^\prime= A_1+\frac{2\pi q}{N}\frac{d\theta}{2\pi}$ for $\partial \sigma=\Sigma$ and modifying the periodicity condition for $\Psi$ so that it obeys
\eq{\label{multivaluedpsi}
\Psi(\theta)=e^{\frac{2\pi i q}{N}}\Psi(\theta+2\pi)
}
where $\theta$ is the polar angle in the $x-y$ plane. This eliminates $\CS_q$ and then integrating out $B_2$ restricts $A_1^\prime$ to be a $\IZ_N$ gauge field while at the same time inducing an Aharonov-Bohm phase on $\Psi$. 

Another way to see that the operator 
$\CS_q$ both turns on the gauge field and causes $\Psi(x)$ to be multi-valued 
is to use the operator formalism of generalized symmetries.  Analyzing the Lagrangian in \eqref{LagFerm} shows that the  gauge field is shifted $A\mapsto A+\alpha \,\delta(\sigma)$ by the operator\footnote{Note that here we are working with singular transformations which are a typical feature of  the operator formalism.}
\eq{
V_\alpha(\Sigma;\sigma)=e^{i \alpha \oint_\Sigma \frac{NB_2}{2\pi}+i \alpha \int_\sigma \ast j_\Psi}
}
where $\partial \sigma=\Sigma$ and the operator $e^{i \alpha \int_\sigma \ast j_\Psi}$ induces a phase jump $e^{i \alpha}$ of  the $\Psi$ field across the manifold $\sigma$. Since $\CS_q(\Sigma)=V_{\frac{2\pi q}{N}}(\Sigma;\sigma)\,  e^{-\frac{2\pi i q}{N}\int_\sigma \ast j_\Psi}$, it is clear that $\CS_q$ both shifts the gauge field and causes $\Psi(x)$ to be multi-valued.




Similarly, a scalar field of charge $Q$ will have an analogous periodicity condition:
\eq{\label{scalarperiod}
\Phi(\theta)=e^{\frac{2\pi i qQ}{N}}\Phi(\theta+2\pi)
}
The consequence of this periodicity condition is that the system is invariant under rotations which are generated by $\hat{L}_z=i D_\theta$.

\vspace{-0.5cm}
\subsection{Scattering Cross Section}
\vspace{-0.4cm}
Consider scattering charge 1 fermions $\Psi$ perpendicularly off of the string along the $x$-direction. Due to the $z$-translation symmetry, the problem reduces to a $(2+1)d$ problem. To compute the scattering cross section, decompose the wave function into an incident ($\Psi^{(i)}$) and a scattered cylindrical wave $(\Psi^{(s)}$)
\eq{\label{incidentscatterdecomp}
\lim_{r\to \infty}\Psi=\Psi^{(i)}+\Psi^{(s)}
}
Importantly, $\Psi^{(i)}$ and $\Psi^{(s)}$ are both multi-valued as in equation \eqref{multivaluedpsi}. 

We need the mode expansion of the fermions in the background with fixed $A_1$ given by 
\eq{
A_1 =\frac{q}{N}d\theta
}
If we expand $\Psi$ in terms of 2-component Weyl fermions $\Psi=(\psi,\bar\chi)^T$ 
and use the $\gamma$-matrix conventions from \cite{Alford:1988sj}
\eq{
\gamma^0 &= \begin{pmatrix}
    \sigma_3 & 0 \\
    0 & -\sigma_3
\end{pmatrix}\quad~,\quad \gamma^1 = \begin{pmatrix}
    i\sigma_2 & 0 \\
    0 & -i\sigma_2
\end{pmatrix}\\
 \gamma^2 &= \begin{pmatrix}
    -i\sigma_1 & 0 \\
    0 & i\sigma_1
\end{pmatrix}~~, \quad \gamma^3 = \begin{pmatrix}
    0 & \mathds{1} \\
    -\mathds{1} & 0
\end{pmatrix}
}
then the equations for $\psi,\chi$ decouple and can be written as  
\eq{
(\omega-m)\psi_1=&-ie^{- i \theta}\left(\partial_r-\frac{i}{r}\left(\partial_\theta+\frac{ i q}{N}\right)\right)\psi_2
\\
(\omega+m)\psi_2=&-ie^{ i \theta}\left(\partial_r+\frac{i}{r}\left(\partial_\theta+\frac{i q}{N}\right)\right)\psi_1
}
where 
\eq{
\psi=e^{-i \omega t}\begin{pmatrix}\psi_1\\\psi_2\end{pmatrix}
}
and similarly for $\chi_{1,2}$ with $q\to -q$. Using the periodicity condition 
\eq{
\psi(\theta)=&e^{2\pi i q/N}\psi(\theta+2\pi) 
\\
 \chi(\theta)=&e^{-2\pi i q/N}\chi(\theta+2\pi)
}
we find that the field can be decomposed into a linear combination of the $\hat{L}_z$ eigen-modes 
\eq{\label{psimodes}
\psi_{n}^{(\pm)}=  \,e^{-i\frac{q}{N}\theta- i \omega t}\begin{pmatrix}
J_{\pm n}(k r) e^{i n \theta}\\
\frac{\pm i k }{\omega+m}J_{\pm (n+1)}(k r)\,e^{i (n+1)\theta}
\end{pmatrix}
}
where $k^2=\omega^2-m^2$. 
Notice that  the $q$-dependence only appears in the periodicity of $\theta$ -- this 
follows from the fractional flux in the $A_1$ gauge field induced by the Aharonov-Bohm string and is required by the corresponding periodicity of $\Psi(x)$ in \eqref{multivaluedpsi}. This expansion differs from the standard expansion of a fermion in cylindrical coordinates only by the overall phase. Crucially, our expansion differs from that of  \cite{Alford:1988sj} which did not obey the periodicity condition imposed by the $\CS_q$ operator. 

Consider scattering an incident wave 
\eq{
\psi^{(i)}=\begin{pmatrix}1&\\\frac{-ik}{\omega+m}\end{pmatrix}e^{-i \omega t- i k x -i\frac{q}{N}\theta}~,
}
which corresponds to incoming fermions along the $x$-direction
\eq{
j^\mu_\Psi=i\bar\Psi\gamma^\mu \Psi=\left(\frac{2im}{\omega+m},\frac{2 k}{\omega+m},0,0\right)~.
}
Using the identity 
\eq{\label{Jidentity}
e^{- i k x-\frac{iq}{N}\theta}=e^{ -\frac{iq}{N}\theta}\sum_{n\in \IZ}i^nJ_n(k r) e^{ i n \theta}~,
}
we can expand the wave function into incident and scattered components as in \eqref{incidentscatterdecomp}. Using the mode expansion in \eqref{psimodes}, we find that $\Psi^{(s)}=0$ and therefore that the differential scattering cross section is zero:
\eq{
\frac{d\sigma}{d\theta}=0
}
\noindent \underline{Comments:} 
\begin{itemize}
\item We only impose that the multi-valued wave function be finite as $r\to 0$ since the cosmic string in the $\IZ_N$ gauge theory is not a disorder operator. 
Additional local interactions on the cosmic string can be incorporated by upgrading the string to a defect operator with additional boundary conditions/interactions on its world volume. 

\item Our computation reproduces the experimentally observed effects of the Aharonov-Bohm effect. Recall that the double-slit experiment requires two wave packets to scatter around the solenoid. The phase dependence we have noted for charged fields will lead to phase differences between the two wave packets and will produce the observed interference effect. Notice however that if $\frac{d\sigma}{d\theta}\neq 0$ for the single particle scattering, there would be an additional observable interference pattern in the single-slit experiment in the presence of a solenoid 
due to the asymptotic decomposition of the wave function 
\eq{
\lim_{r\to \infty}\Psi(x)=e^{-i k x-i \frac{q}{N}\theta}\psi_0+\frac{e^{i kr -i \frac{q}{N}\theta}
}{\sqrt{r}}\psi_1(\theta)}
for spinors $\psi_0$, $\psi_1(\theta)$. 


\item In a UV complete theory where the cosmic string is the IR limit of a smooth field configuration, we expect that there will be a non-vanishing scattering cross section. The vanishing of the scattering cross section in the effective IR theory can be interpreted as the statement that the scattering cross section is suppressed by the core size  $R$ which has been computed for a hard-core string in \cite{Everett:1981nj} to be
\eq{
\sigma_{tot}\sim \frac{\pi^2}{k\,\log^2(k R)}+O(R)
}
so that $\sigma_{tot}\to 0$ as $kR\to 0$. 
\end{itemize}

 We can similarly compute the (transverse) differential scattering cross section of a charge 1 scalar field $\Phi$ off of $\CS_q$:
\eq{
\lim_{x\to \infty}\Phi=\Phi^{(i)}+\Phi^{(s)}
} 
where 
\eq{
 \Phi^{(i)}=e^{-i k x-i\frac{q}{N}\theta}\quad, \quad \Phi^{(s)}=\frac{f(\theta)}{\sqrt{r}} e^{ i k r-i \frac{q}{N}\theta}
}
The mode expansion of $\Phi$ is given by solutions of 
\eq{
\left(\omega^2-m^2+\partial_r^2+\frac{1}{r}\partial_r +\frac{1}{r^2}\left(\partial_\theta+ \frac{ i q}{N}\right)^2\right)\phi=0
}
where $\Phi=e^{- i \omega t}\phi(r,\theta)$. Using the periodicity condition in \eqref{scalarperiod} and assuming regularity of $\Phi$ as $r\to 0$ results in
\eq{
\Phi(x)=e^{- i \omega t-i\frac{q}{N}\theta}\sum_n c_{n}~e^{- i n \theta} J_n(k r)~,
}
where $k^2=\omega^2-m^2$. As before, using the identity \eqref{Jidentity}, we can fully expand the incoming plane wave using $c_{n}=(i)^n$ so that $\Phi^{(s)}=0$ and the differential scattering cross section $\frac{d\sigma}{d\theta}=|f(\theta)|^2$ vanishes up to corrections suppressed by the string core size. 

\vspace{-0.5cm}
\section{Application to Callan-Rubakov Effect for Cosmic Strings}
\vspace{-0.4cm}

Enhanced scattering of fermions off of a cosmic string (such as proposed in  \cite{Alford:1988sj}) can be used to enhance baryon number violating processes in beyond Standard Model theories with cosmic strings as proposed in \cite{Alford:1989ie}.

Consider a theory with a cosmic string coupled to a pair of Dirac fermions $\psi,\chi$ -- where we think of $\psi$ as a quark and $\chi$ as a lepton which individually transform under $U(1)_B$, $U(1)_L$ respectively. 
Let us scatter a $\psi$ off of a cosmic string and decompose the fields into incident $(\psi^{(i)})$ and scattered $(\psi^{(s)},\chi^{(s)})$ wave-functions as before:
\eq{
\lim_{r\to \infty}\begin{pmatrix}\psi\\\chi\end{pmatrix}=\begin{pmatrix}\psi^{(i)}\\0\end{pmatrix}+\begin{pmatrix}\psi^{(s)}\\\chi^{(s)}\end{pmatrix}
}
If there is a scattering effect such that $\psi^{(i)}$ activates a large $\psi^{(s)}$, then any boundary conditions that relate $\psi$ to $\chi$ (which can be activated by additional UV interactions as below) will additionally generate a large scattered $\chi^{(s)}$. Since the scattering $\psi\to \chi$ violates $U(1)_B$, we are lead to a large enhancement of baryon number violation.

A simple  model that  dresses  Aharonov-Bohm cosmic strings with $U(1)_B$-violating boundary conditions \cite{Alford:1988sj} is a $U(1)_g$ gauge theory that couples to $\psi,\chi$ with  charges $\pm1$ respectively, a scalar field $\eta$ with charge $N$, and a neutral scalar field $\phi$. Assume that there is a scalar potential and that the fermions interact via a Yukawa coupling
\eq{
\CL=&\frac{1}{2g^2}F_{\mu\nu}^2+\bar\psi(i \slashed{D}-m_\psi)\psi+\bar\chi(i \slashed{D}-m_\chi)\chi\\
&-|D_\mu\eta|^2-|\partial_\mu\phi|^2
-V(\phi,\eta)
-\lambda \,\bar\phi \psi \chi+c.c. 
}
The global symmetries of the theory are listed in the table below for completeness 
with $\psi,\chi$  decomposed into Weyl components $\psi_\pm,\chi_\pm$ respectively.

\begin{center}
\begin{tabular}{c|ccc}
&$U(1)_g$&$U(1)_{B-L}$&$U(1)_B$\\
\hline
$\psi_+$&$1$&$1$&$1$\\
$\psi_-$&$-1$&$-1$&$-1$\\
$\chi_+$&$-1$&$-1$&0\\
$\chi_-$&$1$&$1$&0\\
\hline
$\eta$&$N$&0&0\\
$\phi$&0&$0$&1
\end{tabular}
\end{center}

In this model, we choose $V(\phi,\eta)$ such that $\eta$ condenses and $\phi$ gains a large mass so that we flow to a $\IZ_N$ gauge theory with cosmic strings that couple to a pair of Dirac fermions of charge 1. We additionally include an interaction $V=...+ m|\eta|^2|\phi|^2$ ($m>0$) in the scalar potential so that $\phi$ is generically massive, but 
 condenses inside the core of the cosmic string. This spontaneously breaks $U(1)_B$  and consequently localizes a Goldstone degree of freedom in the cosmic string core. The 
cosmic string operator is now a defect operator  with  boundary conditions that relate $\psi$ to $\chi$ and preserve $U(1)_{B-L}$ while  breaking $U(1)_B$.\footnote{In the low energy effective theory, the cosmic string operator will require an insertion of $\CS_q$ in addition to  boundary conditions on $\psi,\chi$ on the string world sheet induced by their localized interaction with $\phi$.} The reason why the effective IR boundary conditions break baryon number even though it is preserved by the UV Lagrangian is that the scattering process deposits $U(1)_{B}$ charge into the string core through the localized Goldstone mode.

However, as we have shown, the Aharonov-Bohm interaction does not give rise to a large enhancement (i.e. unsuppressed relative to the core size) 
to the scattering cross section. Therefore, the baryon number violation from scattering off of an Aharonov-Bohm cosmic string with $U(1)_B$ violating boundary conditions as in this model will be suppressed by the geometric size of the cosmic string core. 

In cosmological settings, $U(1)_B$-violating defects generically wash out baryon asymmetry. However, since matter scattering  off of Aharonov-Bohm strings does not have an amplified cross section, we find that it  does not necessarily wipe out large baryon asymmetry for sufficiently small core size. 
But, it has been suggested in \cite{Brandenberger:1988rp,Brandenberger:1988as} that a large proliferation of baryon number violating cosmic strings may still be able to wash out early universe baryon asymmetries. We leave further investigation of this to future studies.

\section*{Acknowledgements}
\vspace{-0.4cm}
The authors would like to thank Ken Intriligator,  Aneesh Manohar, Oren Bergman, Ben Grinstein, Sungwoo Hong, John March-Russell, LianTao Wang, Seth Koren, and Prateek Agrawal for helpful discussions and related collaborations. 
TDB is supported by Simons Foundation award 568420 (Simons
Investigator) and award 888994 (The Simons Collaboration on Global Categorical Symmetries). JSG acknowledges the support of the Natural Sciences and Engineering Research Council of Canada (NSERC) [PGS D - 587640] and U.S. Department of Energy (DOE) award number~DE-SC0009919. JSG also thanks the Galileo Galilei Institute for Theoretical Physics for their hospitality during the completion of this work.


\bibliographystyle{utphys}
\bibliography{CosmicStringScatteringbib}

\end{document}